\documentclass[referee]{raa}
\usepackage{graphicx,times}
\usepackage{natbib}
\usepackage{amssymb,amsmath}
\usepackage{rotating}
\bibpunct{(}{)}{;}{a}{}{,}
\usepackage[a4paper=true,dvipdfm=true,pagebackref=true]{hyperref}
\hypersetup{pdftitle = The title of my PDF, pdfauthor = My name, pdfsubject= The subject, pdfkeywords = keyword1 keyword2 keyword3}
\hypersetup{colorlinks = true, linkcolor = green, anchorcolor = red, citecolor = blue, filecolor = red, pagecolor = red, urlcolor = red}

\begin{document}

   \title{The M-giant star candidates identified in the LAMOST data release 1
$^*$
\footnotetext{\small $*$ Supported by the National Natural Science Foundation of China.}
}

 \volnopage{ {\bf 2015} Vol.\ {\bf X} No. {\bf XX}, 000--000}
   \setcounter{page}{1}

   \author{Jing Zhong\inst{1},  S\'ebastien L\'epine\inst{2,3,4}, Jing Li \inst{1}, Li Chen \inst{1}, Jinliang Hou, \inst{1}, Ming Yang, \inst{5}, Guangwei Li, \inst{5}, Yong Zhang,\inst{6}, Yonghui Hou, \inst{6} }

   \institute{Key Laboratory for Research in Galaxies and Cosmology,
Shanghai Astronomical Observatory, Chinese Academy of Sciences,
80 Nandan Road, Shanghai, China; {\it jzhong@shao.ac.cn}\\
   \and
   Department of Physics \& Astronomy, Georgia State
  University, 25 Park Place, Atlanta, GA 30303, USA;
   \and
   Department of Astrophysics, Division of Physical
  Sciences, American Museum of Natural History, Central Park West at
  79th Street, New York, NY 10024, USA;
  \and
  City University of New York, The Graduate Center, 365
  Fifth Avenue, New York, NY 10016, USA;
  \and
  Key Laboratory of Optical Astronomy,
  National Astronomical Observatories, Chinese Academy of Sciences,
  Beijing 100012, China
  \and
  Nanjing Institute of Astronomical Optics \& Technology,
  National Astronomical Observatories, Chinese Academy of Sciences,
  Nanjing 210042, China
}

\abstract{
We perform a discrimination procedure with the spectral index diagram of TiO5 and CaH2+CaH3 to separate M giants from M dwarfs. Using the M giant spectra identified from the LAMOST DR1 with high signal-to-noise ratio (SNR), we have successfully assembled a set of M giant templates, which show more reliable spectral features.Combining with the M dwarf/subdwarf templates in \cite{zhong.2015}, we present an extended M-type templates library which includes not only M dwarfs with well-defined temperature and metallicity grid but also M giants with subtype from M0 to M6. Then, the template-fit algorithm were used to automatically identify and classify M giant stars from the LAMOST DR1. The result of M giant stars catalog is cross-matched with 2MASS JHK$_{s}$ and WISE W1/W2 infrared photometry. In addition, we calculated the heliocentric radial velocity of all M giant stars by using the cross-correlation method with the template spectrum in a zero-velocity rest frame. Using the relationship between the absolute infrared magnitude M$_{\rm J}$ and our classified spectroscopic subtype, we derived the spectroscopic distance of M giants with uncertainties of about 40\%.  A catalog of 8639 M giants is provided. As an additional search result, we also present 101690 M dwarfs/subdwarfs catalog which were classified by our classification pipeline.
}

   \authorrunning{J. Zhong et al. }            %author_head in even pages
   \titlerunning{The Catalog of M-type stars}  % title_head in odd pages
   \maketitle

%________________________________________________ sections below
%
\section{Introduction}           %% first-level sections will be auto-capitalized
\label{sect:intro}

M giants are red-giant-branch (RGB) stars with low surface temperature ( $<$ 4000 K) and high luminosity (log $L/L_{\sun}$ $\sim$ 3-4) in the late-phase of stellar evolution.  Its luminous nature allows us to use these stars as good tracers to study the outer Galactic halo and distant substructures. By selecting the M giant candidates from the Two Micron All Sky Survey (2MASS), \cite{2003ApJ...599.1082M} mapped out the first global view of Sagittarius dwarf galaxy all over the sky and found that a significant fraction of halo M giants in the Milky Way were contributed by the Sagittarius Dwarf Galaxy; \cite{2010ApJ...722..750S} identified 16 candidate stellar halo structures at high Galactic latitude, of which 6 are new. To explore the distant region of our Galaxy's outer halo, 404 M giants were identified from the UKIRT Infrared Deep Sky Survey \citep[UKIDSS;][]{2007MNRAS.379.1599L}, and the kinematic analysis indicated that the M giant candidates can be used to constrain the number of Sagittarius accretion event \citep{2014AJ....147...76B}. Moreover, two extremely distant giants have been confirmed by spectroscopy with distance of $\sim$ 240-270 kpc, almost beyond the virial radius of our Galaxy \citep{2014ApJ...790L...5B}.

Until currently, most of M giant candidates are selected from the photometric database, only a small fraction of them obtained via visible/infrared spectra \citep{1994A&AS..105..311F,1994PASP..106..382D,1995AJ....109.1379A,1999ApJS..123..283M,2000A&AS..146..217L,2012ApJ...753...90M}. In the Sloan Digital Sky Survey \citep[SDSS;][]{2000AJ....120.1579Y}, the M giants fraction in the M-type spectroscopic sample is about 0.5$\%$-1.0$\%$ \citep{2011AJ....141...97W,2008AJ....136.1778C}, corresponding to several hundred giant spectra. As an alternative efficient spectroscopic survey, the LAMOST Galactic survey project observed more M giant candidates than SDSS. In the LAMOST pilot survey, \cite{2014AJ....147...33Y} present 58,360 M dwarf candidates and estimate the M giants contamination as about 4$\%$ by using the J-H color criteria \citep{1988PASP..100.1134B}. Comparing with the SDSS survey, the high spectrum acquiring rate and high giant fraction observing rate indicate the great potential of LAMOST survey program to establish the largest M giant spectroscopic sample for future research.

In \citet [here after Z15]{zhong.2015}, we have performed a spectral classification of all M dwarf stars in the LAMOST commissioning data. Using the template-fit method, 2612 spectra with relatively high SNR were positively identified as M dwarf spectra. By examining some outliers in the spectral index distribution, we found a few spectra in our sample are more likely belong to M giants instead of M dwarfs. As we pointed out in the Z15, the giants misclassification are mainly caused by the lack of giant templates in our automated classification pipeline. Although the giants contamination is not significant in the commissioning survey, it is necessary to fix the shortcoming in our pipeline since the giants fraction is largely increased in the pilot survey and regular survey. Our effort of assembling the M giant templates will be devoted to well classify the M-type stars including the M dwarfs/subdwarfs and M giants.

In this paper, a brief description of the LAMOST DR1 is given in section~\ref{data}. In section~\ref{templates}, we mainly introduce our effort to establish the M giant spectral templates, including the luminosity discrimination and the temperature classification. Combining with the M giant and dwarf templates, we used the revised classification pipeline to classify M giant stars with different subtype in LAMOST DR1, the analyses and results of our classification are presented in section~\ref{sect:result}. In the last section, a brief conclusion and discussion is provided.

%=...=%%%%%%%%%%%%%%%%%%%%%%%%%%%%%%%%%%%%%%%%%%%%%%%%%%%%%%%%%%%
\section{The LAMOST DR1 data}
\label{data}

The LAMOST survey is a spectroscopic survey for stars and extra-galaxies. Based on the quasi-meridian reflecting Schmidt telescope\citep{Cui.2012} of 4 meters effective aperture and 4000 optical fibers, the LAMOST survey become a most ambitious spectroscopic observation program to acquire over 10 million spectra of stars and galaxies in five years \citep{2012RAA....12..723Z}. At present, the LAMOST survey has gone through the commissioning phase (2009-2011), the pilot survey (2011-2012), the first year (2012-2013) and the second year (2013-2014) regular survey.

The LAMOST Data Release One (DR1) includes the pilot survey and the first year regular survey data, with totally of 2,204,860 spectra\citep{luo15a}. Most of these spectra are observed with an implementing of the LAMOST Experiment for Galactic Understanding and Exploration (LEGUE) survey \citep{deng12}. For all 1,944,406 stellar spectra in DR1, the SNR are greater than 10 in the SDSS g or r band. A subset of about 1.1 million stellar spectra (AFGK stars) with relatively high SNR provides the stellar parameters like the effective temperature (T$_{\rm eff}$), surface gravity (Logg), metallicity (Fe/H]) and radial velocity (RV).

The LAMOST spectra are first reduced by the LAMOST 2D pipeline over the vacuum-wavelength scale from 3800 $\AA$ to 9000 $\AA$, which mainly include the processes of bias subtraction, flat correction, skyline substraction, wavelength calibration and flux calibration \citep{luo12}. The extracted spectra are then passing through the 1D pipeline to classifying the spectral type and calculating the radial velocity and redshift\citep{luo15b}.

%=...=%%%%%%%%%%%%%%%%%%%%%%%%%%%%%%%%%%%%%%%%%%%%%%%%%%%%%%%%%%%
\section{ M-giant spectral templates }
\label{templates}
\subsection{Luminosity class}

To determine the luminosity class of M type stars, specialized discrimination-methods have been developed over the years which are based on the colors, proper motions and spectral indices. The color discrimination-method was first introduced by \cite{1988PASP..100.1134B}. They show that M giants and M dwarfs distribute on the different locus in the [J-H, H-K] color-color diagram, which are mainly caused by the opacity differences of molecular bands of H$_2$O \citep{1998A&A...333..231B}. Since M giants and M dwarfs occupied distinct loci, with the giants group having relatively large distant and small proper motion, \cite{2011AJ....142..138L} developed a robust method using reduced proper motions ( H$_{\rm V}$) to separate the two luminosity classes of M type stars. According to the comparison of M giant and M dwarf spectra, \cite{2012ApJ...753...90M} suggested a spectroscopic luminosity class algorithm using several gravity-sensitive molecular and atomic spectral indices. In addition, the Mg$_{\rm 2}$ versus $g-r$ was also used as an effective method for discrimination \citep{2008AJ....136.1778C}.

Since the surface gravity are totally different for giant and dwarf, one can use the spectral features as gravitational indicators to determine the luminosity class. For late-type stars, the comparison of giant and dwarf spectra with similar effective temperature shows that at least six molecular and atomic spectral indices in the optical wavelength bands are sensitive to gravity \citep{2012ApJ...753...90M}, such as Na {\sc i} (5868-5918 \AA), Ba {\sc ii}/Fe {\sc i}/Mn {\sc i}/Ti {\sc i} (6470-6530 \AA), CaH2 (6814-6846 \AA), CaH3 (6960-6990 \AA), TiO5 (7126-7135 \AA), K {\sc i} (7669-7705 \AA), Na {\sc i} (8172-8197 \AA), Ca {\sc ii} (8484-8662 \AA). In our work, considering the narrow wavelength region for atomic spectral indices and possible skylines contamination in red region around 8000 \AA, the molecular spectral indices of TiO and CaH were used to separate M giants from dwarfs in the LAMOST DR1 data.

\begin{figure}[t]
\begin{center}
\vspace{0cm}
\hspace{0cm}
\includegraphics[angle=0,scale=0.75]{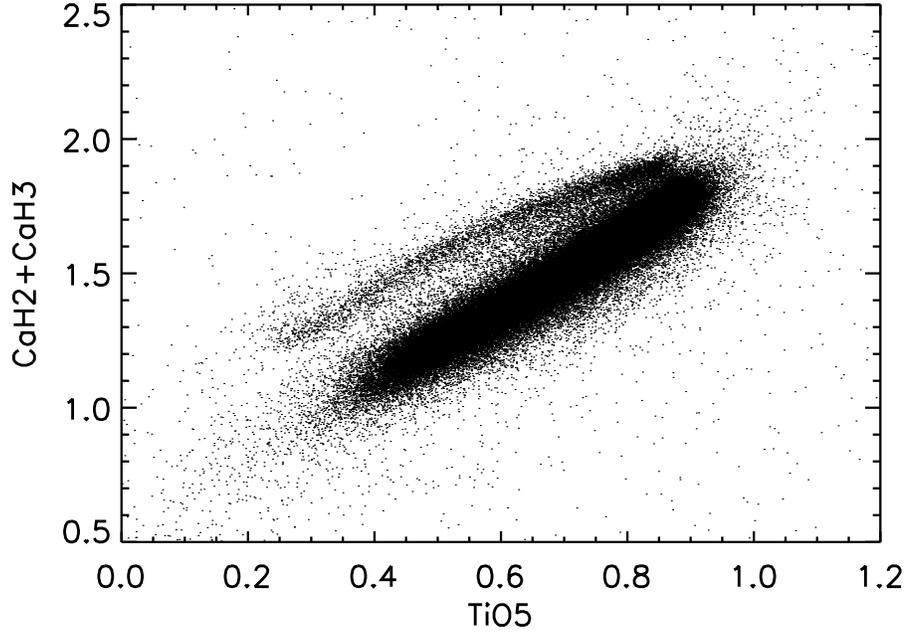}
\caption{
The M type stars distribution in the CaH2+CaH3 against TiO5 diagram. Two branches in this diagram clearly indicate the two populations. Because of the weaker CaH molecular bands, about 10000 M giants are located in the upper branch. Comparison with previous results \citep{2007ApJ...669.1235L}, the stars distribute in the lower branch are mainly M dwarfs/subdwarfs. The clear separation of different population in this diagram indicate the great potential of using spectral indices to distinguish M giants and M dwarfs.
\label{idx}}
\end{center}
\end{figure}

First, we used the template-fit method (Z15) to select M-type spectra which positively present the characteristic molecular features, e.g., TiO, VO and CaH. Then the spectral indices of TiO5, CaH2 and CaH3, as defined by \cite{1995AJ....110.1838R} and \cite{2007ApJ...669.1235L}, were calculated. Figure~\ref{idx} shows the spectral indices diagram for all M type stars we identified in the LAMOST DR1. Two populations are clearly distinguishable in this spectral indices diagram. Giants with weaker CaH molecular bands are located on the upper branch, which are consistent with the giant/dwarf discrimination by \cite{2012ApJ...753...90M}. The number of giant candidates in the upper branch are about 10,000.

\subsection{Temperature type}

As shown in \cite{2007AJ....134.2398C}, the SDSS $r-i$ color for late-type stars has shown good relationship with the Morgan-Keenan (MK) spectral subtypes, which spans about 2 mag from M0 to M10. To provide spectral subtypes along the temperature sequence for M giants, we choose the SDSS $r-i$ color as an indicator to classify M giant subtypes.

In order to select high quality LAMOST spectra as training spectra for each spectral subtype grid, we first cross-matched the giant candidates with SDSS DR9 photometric database. Because a large number of LAMOST stars are located in the Galactic anti-center region, only about 3600 candidates have the SDSS $ugriz$ photometry information. Next, to reduce the extinction effect and to select reliable photometry, a giant candidate has to meet the following criteria:

(1) The $r$ band extinction on the Schlegel's Galactic extinction map \citep{1998ApJ...500..525S} must be less than 0.2.

(2) The \texttt{fphotoflags} in SDSS photometry must include BRIGHT flag=0, EDGE flag=0, (BLENDED flag ${\&}$ NODEBLEND flag)=0, COSMIC$\_$RAY flag=0 and SATURATED flag=0.

(3) The $g-r$, $r-i$ color band must distribute on the M type star's locus with 1.0 $< g-r <$ 1.4 mag and 0.5 $< r-i <$ 2.8 mag.

Upon these criteria, the training sample was cut down from $\sim$ 3600 to $\sim$ 600. Then the remaining giant candidates were confirmed by manual inspection. Giant spectra which are suffer from sky lines contaminations, serious reddening, low signal-to-noise ratio or displaying the characteristic of binary spectrum, were excluded from the training sample. Finally, approximately 200 high quality giant spectra with good photometry in SDSS were left as the training spectra to assemble a grid of temperature sequence.

Table~\ref{table1} list the $r-i$ color ranges for MK spectral subtype grid, which mainly based on \cite{2007AJ....134.2398C}. Since in our sample there is no giant candidate with $r-i$ color greater than 2.0 mag, the synthetic M giant templates span the spectral subtypes from M0 to M6. For spectra with overlapping $r-i$ colors between two spectral type bins, we manually assigned the spectra by eyes and make sure that the difference in spectral type is $\pm$ 1 subtype.

\begin{table*}

\caption{The Color Ranges of subtype classification }
 \label{table1}
\begin{tabular}{rc}
\hline
 Spectral Type  &     $ r-i$\\
\hline
   M0.............. & 0.50-0.65  \\
   M1.............. & 0.58-0.80  \\
   M2.............. & 0.70-0.95  \\
   M3.............. & 0.90-1.10  \\
   M4.............. & 1.00-1.35  \\
   M5.............. & 1.30-1.70  \\
   M6.............. & 1.60-2.00 \\
   \hline
   \end{tabular}
\end{table*}

\subsection{Radial velocity correction}

To correct the radial velocity for each training spectrum, we manually used the IRAF/\texttt{rv.rvidlines} package to measure the wavelength correction to the zero-velocity rest-frame. Since most of atomic lines in the optical band are weak in the M giant spectrum, the near-infrared calcium (Ca {\sc ii}) triplet at 8498, 8542, and 8662  {\AA} were predominantly measured as reference-wavelength. In addition, we also used H$_\alpha$ (6563 \AA) absorption line to calibrate early type M giants (earlier than M4), which display significant H$_\alpha$ absorption feature. After measuring the wavelength correction, the training spectra were shifted toward blue or red to the zero-velocity rest frame according to their correction. The maximum radial velocity correction in our training sample is approximately equal to $\pm $200 km/s. Then the corrected spectra were measured and shifted again. This procedure was repeated until the measured radial velocity for each corrected training spectrum is less than 5 km s$^{-1}$.

\subsection{Template spectra}

\begin{figure*}[t]
\begin{center}
\vspace{0cm}
\hspace{0cm}
\includegraphics[angle=0,scale=0.8]{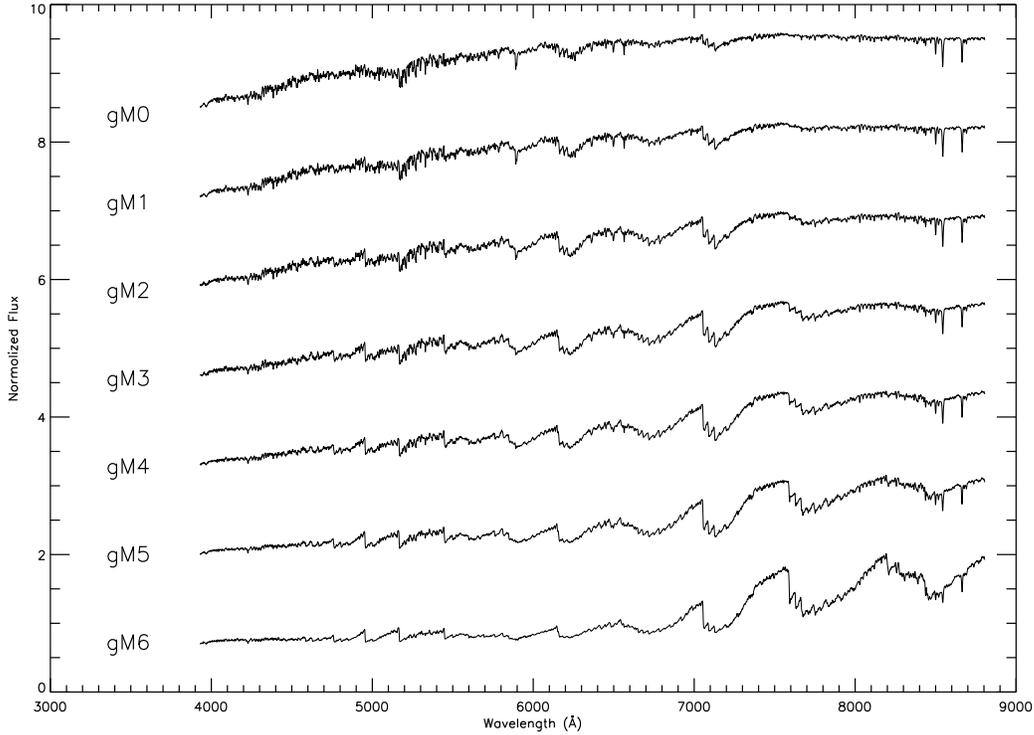}
\caption{
The M giant templates from M0 to M6. We defined seven different giant subtypes based on the r-i colors, as proposed in Table~\ref{table1}. Each template spectrum is assembled from at least five LAMOST high SNR spectra which are confirmed by manual assignment. From top to bottom, the increasing strength of molecular bands, such as CaH, TiO and VO, reflect the decreasing temperature of giant spectra. The template spectra in this figure can be retrieved from the Strasbourg astronomical Data Center (CDS).
\label{tmp}}
\end{center}
\end{figure*}

The wavelength corrected spectra were used to assemble the template spectra before normalizing at 8350 {\AA} \citep{2007AJ....133..531B}. For each spectral subtype bin, at least five training spectra were combined to create the synthetic template spectra.

Figure~\ref{tmp} present the M giant template spectra from M0 to M6, which were assembled by the LAMOST DR1 spectra. From top to bottom, the spectra were presented along their temperature sequence.

\begin{figure}[t]
\begin{center}
\vspace{0cm}
\hspace{0cm}
\includegraphics[angle=0,scale=0.75]{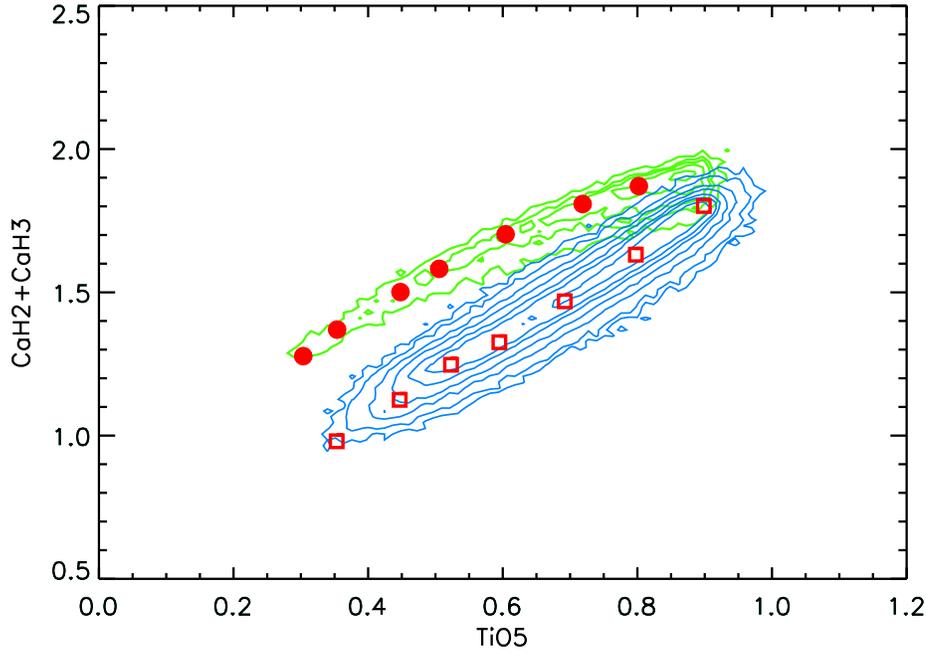}
\caption{
To verify the reliability of temperature sequence in our template classification, we add the spectral subtypes of giant templates into the spectral indices diagram. From right to left, seven red dots represent the seven M giant templates from M0 to M6. The distribution indicate that our synthetic templates define a reliable temperature grid.
\label{tmpidx}}
\end{center}
\end{figure}

To verify the reliability of our subtype classification, we calculate sets of molecular spectral indices in the synthetic template spectra as giants temperature indicators. Figure~\ref{tmpidx} shows the variations of different spectral indices as a function of our subtype classification. The M giants and dwarfs distribution shown in Figure~\ref{idx} were represented as green contour and blue contour, respectively. The indices of M giant templates are shown as red dots. From right to left, the template subtypes are from M0 to M6, which means the CaH and TiO molecular absorption bands of late-type template are stronger than the early-type template. As a comparison, we also plot the M dwarf templates indices distribution, which are shown as red squares. The spectral indices distribution of giant templates shows that our template spectra are consist with the M giants branch, and also define a reliable temperature grid.

\begin{figure}[t]
\begin{center}
\vspace{0cm}
\hspace{0cm}
\includegraphics[angle=0,scale=0.65]{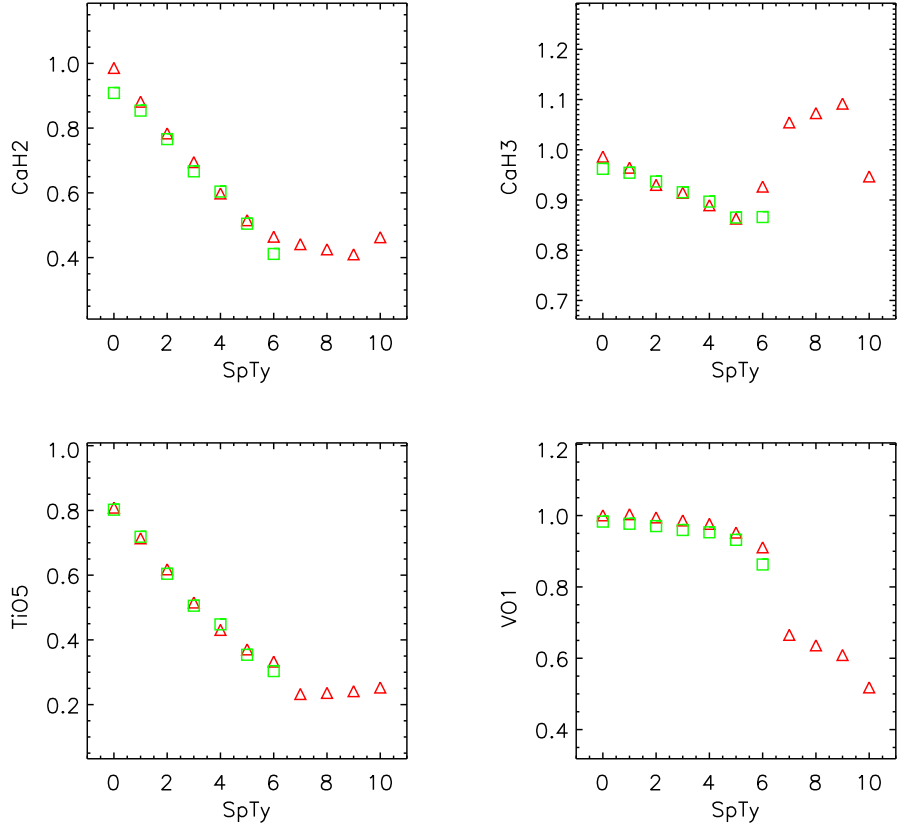}
\caption{
Validation of four spectral indices distribution for the two M giant templates. The red triangles represent the spectral indices of \cite{1994A&AS..105..311F}, with subtype range from M0 to M10. The green squares represent the spectral indices of our templates, from M0 to M6. The similar distribution shows the two M giant templates both defined a reliable spectral subtype grid in early type.
\label{fluksidx}}
\end{center}
\end{figure}

In particular, we compare our M giant templates with Fluks's templates \citep{1994A&AS..105..311F}. Figure~\ref{fluksidx} shows the comparison results of four spectral indices, including CaH2, CaH3, TiO5 and VO1, which were defined in \citet{2013AJ....145..102L}. The templates of \cite{1994A&AS..105..311F} were shown as red dots. These intrinsic spectra were derived from 97 very bright M giant stars in the solar neighbourhood, with spectral subtype range from M0 to M10 and wavelength range from 3800 $\AA$ to 9000 $\AA$. Our templates were shown as green squares. The consistency of spectral indices for early type templates (M0-M5) also indicate the reliability of our classification grid. For late type spectrum M6, there are relatively large difference between two templates. We choose to adopt our template which come from observational spectra assembling rather than interpolated spectra in \citet{1994A&AS..105..311F}.

\section{Spectral analyses and classification results}
\label{sect:result}

\subsection{Spectral classification}

In our previous work (Z15), a set of M dwarf templates have been developed as references for automatically identifying and classifying the M dwarfs in the LAMOST spectroscopic data. Our M dwarf templates were assembled from the M dwarf catalog in the SDSS DR7 \citep{2011AJ....141...97W}. Based on the spectral index method \citep{Lepine.2003,Lepine.2007}, we re-classified the M dwarfs into a tentative temperature-metallicity grid with over 18 elements resolution in temperature (K7.0-M8.5) and 12 elements resolution in metallicity (dMr-usdMp). With these well defined M dwarf templates, the template-fit method was used to determine the spectral type of LAMOST stars.

As we described in Z15, although our M dwarf templates provide more reliable estimate of spectral classification by using the template-fit method, because of lacking the M giant templates in our templates library, a fraction of M giants are mis-classified as M dwarfs. To solve this problem, we created a new M-type spectral templates library by combining the M dwarf/subdwarf templates in Z15 and the M giants templates we described above. In the whole M type templates, there are M dwarf templates with temperature from K7.0 to M8.5 and metallicity from dMr to usdMp, and the M giant templates from M0 to M6. The total number of M type templates is 223.

Based on the M-type templates, we re-run our spectral classification pipeline (Z15) to automatically identify and classify M-type stars with spectra from the LAMOST DR1 data source. In order to avoid the additional reddening effect, both template spectra and the LAMOST spectrum were flux-normalized by a pseudo-continuum (See more details in Z15). In the classification pipeline, the template-fit method is used by calculating the chi-square values between the LAMOST spectrum and each of the template spectra. Then, the template spectrum which has the minimum chi-square value is considered as the best-fit, and its spectral subtype is used to mark the corresponding LAMOST spectrum.

After passing through the 2,204,696 LAMOST DR1 spectra to our spectral classification pipeline, we identified 8639 M giants and 101690 M dwarfs/subdwarfs. The excluded spectra were marked as non-M type spectra of which most are earlier type objects like AFGK stars, and a small fraction of spectra were too noisy to be classified.

\subsection{Radial velocity}

To calculate the radial velocities of all M-type stars in our sample by using the template spectra, the radial velocity correction was applied to shift the templates into a zero-velocity rest frame as much as possible. For the M dwarf templates, the red lines of K {\sc i} doublet (7667 {\AA} and 7701 {\AA}) and Na {\sc i} doublet (8185 {\AA} and 8197 {\AA}) were measured. For the M giant templates, we mainly use the Ca {\sc ii} triplet lines (8498, 8542, and 8662  {\AA}) as reference for correction (see Section~\ref{templates} for more details). For each template spectrum, the corrected radial velocity is less than 5 km s$^{-1}$, which is small enough to be considered as the zero-velocity for low resolution spectrum.

After justifying the template spectra to the rest frame, the cross-correlation method were used to calculate the radial velocity of each M giant spectrum in LAMOST DR1. Since the characteristic molecular bands (TiO, CaH, VO) and atomic lines (K{\sc i}, Na{\sc i}, Ca{\sc ii}) of M-type stars are mainly distributed in the red part, the rectification area for normalization and cross-correlation is between 6800 {\AA} and 8800 {\AA}, which covers most of characteristic wavelength range in M-type stars. Then, the best-fit template which is determined by the classification pipeline was used to calculate the radial velocity of LAMOST stars.

\begin{figure}[t]
   \centering
   \includegraphics[width=14.0cm]{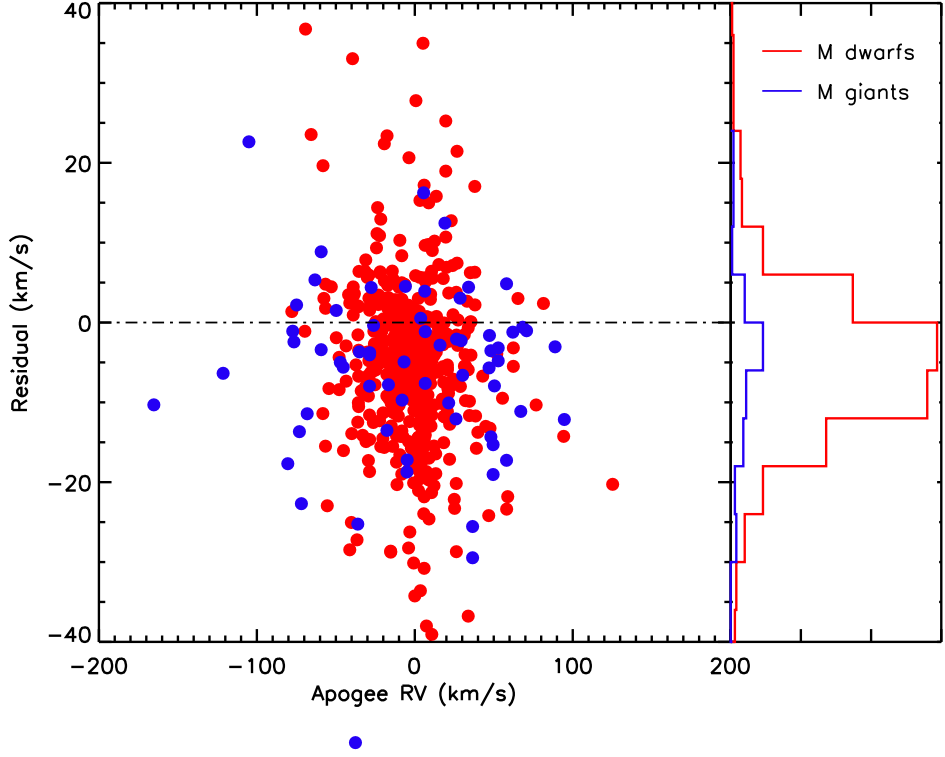}
   \caption{ Distribution of RV residual (RV$_{LAMOST}$-RV$_{APOGEE}$) for all common stars with the LAMOST DR1 and the APOGEE data. The left plot shows the RV residual distribution of 59 M giants (blue dots) and 416 M dwarfs(red dots). The right plot shows the residual histogram distribution of M giants ( blue lines) and M dwarfs (red lines). All the LAMOST spectra we measured have the SNR greater than 10. The $\sigma$ of RV residual in our measurement are 8.4 km s$^{-1}$ for giants and 6.9 km s$^{-1}$ for dwarfs. }
   \label{rv}
\end{figure}

To verify the reliability of our radial velocity measurement, we cross-matched our M-type stars catalog with the APOGEE stellar parameters catalog in DR10. The radial velocity uncertainty in APOGEE data are less than 100 m s$^{-1}$, which can be considered as standard values. Most of stars in LAMOST DR1 are located at the Galactic Anti-center, there are about 67 giants and 575 dwarfs matched in the LAMOST DR1. We exclude common stars which have low SNR in the LAMOST spectra ( SNR $<$ 10 ) or very strange outlier values ( the total numbers are less than 20 ). Finally, we calculate the RV residual values ( RV$_{LAMOST}$-RV$_{APOGEE}$) of 59 common giants and 416 common dwarfs. The RV residual distribution of M giants and M dwarfs are shown in Figure~\ref{rv}. The mean and standard deviation of RV errors in Gaussian fitting are -5.0 $\pm$ 8.4 km s$^{-1}$ for giants and-5.3 $\pm$ 6.9 km s$^{-1}$ for dwarfs. When the SNR criteria of common stars is increased to 30, which reduce the number of giants to 39 and dwarfs to 157, we find that the mean and standard deviation of RV errors are -4.4 $\pm$ 8.1 km s$^{-1}$ and -5.8 $\pm$ 6.5 km s$^{-1}$, corresponding to the M giants and M dwarfs respectively.

\subsection{Estimation of the spectroscopic distance}

To estimate the distances of M-type stars in our sample, we mainly used two independent relationship between the absolute infrared magnitude (M$_{\rm J}$) and the spectroscopic type (SpTy). For M dwarfs, the relationship function was derived by nearby M dwarfs with both spectral types and parallax measurements (Z15). Since most of M dwarfs in our sample are distributed in the solar vicinity, their extinction correction are negligible in the near infrared J band. For M giants, the relationship was referred to the flux calibration and absolute magnitude calculation in the 2MASS system \citep{2007AJ....134.2398C}. Considering that most of M giants are distributed in the distant region, the M giants sample are corrected for extinction using the dust map from  \citet{1998ApJ...500..525S} and the extinction law from Li et al.(2015, in preparation). This extinction law suggest that the extinction coefficients are correlated with Galactic latitude, which is believed to be more reliable in the extinction calculation, especially in the low Galactic latitude. The extinction coefficients toward the Galactic Anti-center are provided in the Appendix of Li et al.(2015, in preparation). Considering the magnitude and extinction uncertainties of 2MASS, the distance accuracy of M dwarfs/giants sample were estimated as 40\%.

\begin{figure}[t]
\begin{center}
\vspace{0cm}
\hspace{0cm}
\includegraphics[angle=-90,scale=0.75]{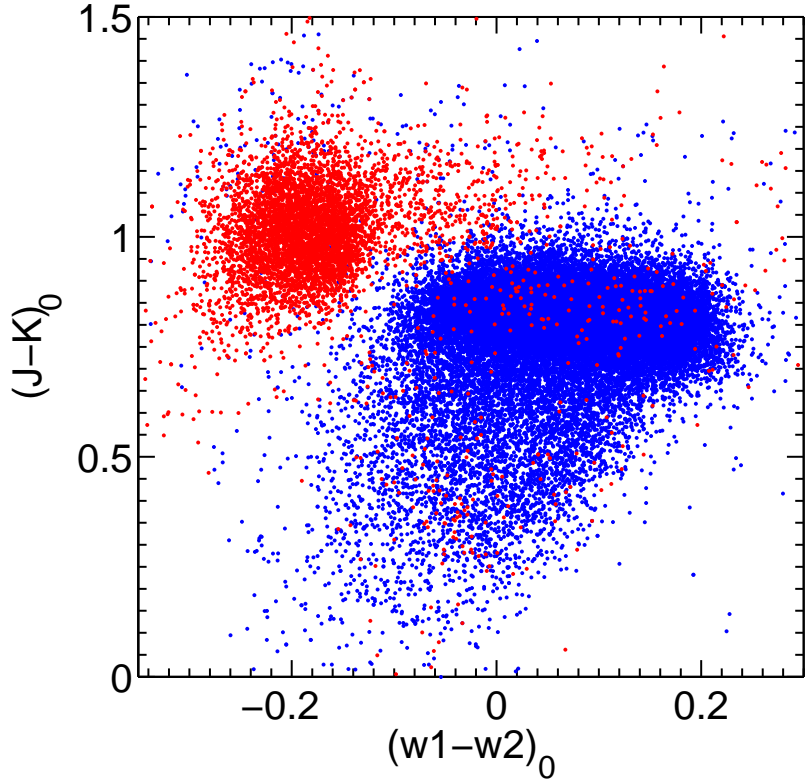}
\caption{
The infrared color distribution of M-type stars. The red dots are M giants and the blue dots are M dwarfs, of which both were classified by our classification pipeline. As expected, the different locations of giants and dwarfs clearly show that our classification pipeline can well separate out the M type stars in the different luminosity. The dwarfs contamination in the  M giants sample is about 4.7\%.
\label{color}}
\end{center}
\end{figure}

\subsection{Catalog description}

In our M-type stars catalog, the proper motion came from the cross-matching with PPMXL catalog \citep{Roeser.2010}, the infrared photometric information are from the 2MASS catalog \citep{Skrutskie.2006} for JHK$_s$ band and WISE catalog \citep{2010AJ....140.1868W} for W1 and W2 band. The M giant and M dwarf candidates are listed in Table ~\ref{tab2}, each of them with ten targets as examples. The completed M giant and M dwarf stars catalog are provided in the electronic version, including the designation in the LAMOST DR1 catalog, the celestial coordinates in epoch 2000, the proper motions and their measurement errors, the infrared photometric magnitudes, the radial velocity measured by our spectral template, the spectroscopic distance we estimated and the spectral subtypes which were classified by our classification pipeline.

Based on our classification, M giants are classified along the temperature sequence, labeled as [gM0.0, gM1.0, gM2.0, gM3.0, gM4.0, gM5.0, gM6.0]. Following the M dwarfs classification in Z15, M dwarfs are classified in the temperature-metallicity grid, which shows [dMr, dMs, dMp, sdMr, sdMs, sdMp, esdMr, esdMs, esdMp, usdMr, usdMs, usdMp] in metallicity and [ K7.0, K7.5, M0.0, M0.5, M1.0, M1.5, M2.0, M2.5, M3.0, M3.5, M4.0, M4.5, M5.0, M5.5, M6.0, M6.5, M7.0, M7.5, M8.0, M8.5] in temperature ( See more details in Z15).

\begin{table*}
%\begin{sidewaystable}
 \centering
\caption{M-type stars catalog with astronomy, photometry, radial velocity, spectroscopic distances and estimated subtypes}
\label{tab2}
\tiny
%\begin{tabular}{p{1.9cm} p{0.9cm} p{0.9cm} p{0.8cm} p{0.6cm} p{0.5cm} p{0.4cm} p{0.55cm} p{0.5cm} p{0.5cm} p{0.5cm} p{0.5cm} p{0.5cm} p{0.5cm} p{0.5cm} p{0.5cm} p{0.45cm} p{0.5cm} p{0.5cm} p{0.5cm} p{0.5cm} }
%\begin{tabular}{p{1.6cm} p{0.8cm} p{0.8cm} p{1.cm} p{1.0cm}  p{0.3cm} p{0.3cm} p{0.3cm} p{0.3cm} p{0.3cm}p{0.3cm}p{0.3cm}p{0.3cm}p{0.5cm}p{0.55cm}p{0.55cm}p{0.65cm}}
\begin{tabular}{ccccccccccccc}
\hline
 Designation & Ra & Dec &  $\mu_{\alpha} \cos(\delta)$ & $\mu_{\delta}$ & J  & H & K$_s$ & W1 & W2 & RV & Dist & SpTy \\
 &  deg & deg & mas yr$^{-1}$ & mas yr$^{-1}$ & mag & mag& mag & mag& mag & km s$^{-1}$ & kpc &   \\
\hline
  J040505.40+285943.6   &      61.27254   &      28.995465  &      4.0   $\pm$     5.0  &  -6.7  $\pm$      5.0  &   10.488  &    9.479  &    9.230  &    9.118  &    9.243  &      -4.3   &   5.16  &  gM0.0\\
 J040611.64+261916.6    &      61.54850   &      26.321288  &     -2.7   $\pm$     4.4  &  -4.9  $\pm$      4.4  &    9.529  &    8.624  &    8.382  &    8.279  &    8.441  &      28.4   &   3.54  &   gM0.0\\
 J041023.67+272143.6    &      62.59864   &      27.362117  &      0.0   $\pm$     5.2  &  -2.9  $\pm$      5.2  &   10.595  &    9.734  &    9.480  &    9.358  &    9.479  &     -19.6   &   5.30  &   gM0.0\\
 J040325.49+293108.0    &      60.85623   &      29.518914  &     -0.6   $\pm$     5.2  &  -4.9  $\pm$      5.2  &    8.485  &    7.501  &    7.141  &    6.985  &    7.169  &      28.4   &   1.83  &   gM5.0\\
 J040329.01+263653.4    &      60.87091   &      26.614842  &     -7.7   $\pm$     4.4  &  -4.9  $\pm$      4.4  &    8.661  &    7.676  &    7.316  &    7.197  &    7.330  &      60.4   &   1.95  &   gM5.0\\
 J070225.22+282327.1    &     105.60509   &      28.390873  &      5.7   $\pm$     5.1  &  -4.7  $\pm$      5.1  &    9.909  &    8.928  &    8.656  &    8.535  &    8.615  &      22.1   &   3.60  &   gM4.0\\
 J065424.57+303015.0    &     103.60241   &      30.504168  &     -5.9   $\pm$     4.1  &   1.4  $\pm$      4.1  &   11.229  &   10.433  &   10.197  &   10.126  &   10.246  &       4.0   &   7.71  &   gM0.0\\
 J065153.21+290913.0    &     102.97172   &      29.153617  &     -0.7   $\pm$     4.9  &  -0.2  $\pm$      4.9  &   10.897  &   10.065  &    9.856  &    9.778  &    9.911  &       5.9   &   6.89  &   gM0.0\\
 J065849.31+303318.9    &     104.70546   &      30.555272  &      3.7   $\pm$     5.1  &  -6.5  $\pm$      5.1  &   11.062  &   10.197  &    9.972  &    9.891  &   10.030  &      19.1   &   7.32  &   gM1.0\\
 J065708.89+303001.0    &     104.28708   &      30.500294  &      0.3   $\pm$     5.0  &  -9.4  $\pm$      5.0  &   10.900  &   10.063  &    9.830  &    9.722  &    9.850  &      13.5   &   6.42  &   gM0.0\\
 J072547.23+300200.1     &    111.44681    &     30.033388   &    -2.3  $\pm$      3.9   &   -12.9    $\pm$    3.9  &   15.062  &   14.371  &   14.251  &   14.140  &   14.122  &      -8.4   &   0.78  &   dMr0.0   \\
 J072602.89+303838.5     &    111.51205    &     30.644054   &     9.4  $\pm$      3.8   &   -22.1    $\pm$    3.8  &   14.648  &   14.013  &   13.732  &   13.682  &   13.586  &      -9.9   &   0.37  &   dMp1.5   \\
 J072724.80+305120.1     &    111.85334    &     30.855586   &    18.8  $\pm$      3.9   &   -12.5    $\pm$    3.9  &   14.817  &   14.227  &   14.125  &   13.963  &   13.954  &     -98.9   &   0.58  &   dKp7.5   \\
 J072650.49+293305.8     &    111.71040    &     29.551627   &    -7.1  $\pm$      3.8   &   -18.7    $\pm$    3.8  &   14.473  &   13.830  &   13.687  &   13.554  &   13.526  &      56.4   &   0.45  &   sdMr0.0  \\
 J072621.99+302531.5     &    111.59163    &     30.425433   &     0.7  $\pm$      3.9   &    -4.9    $\pm$    3.9  &   14.976  &   14.332  &   14.278  &   14.062  &   14.142  &      18.2   &   0.75  &   sdKr7.5  \\
 J071731.61+322753.9     &    109.38173    &     32.464975   &    -8.0  $\pm$      4.0   &   -29.5    $\pm$    4.0  &   12.833  &   12.206  &   11.927  &   11.827  &   11.683  &     -86.2   &   0.10  &   dMr4.0   \\
 J072423.08+285503.2     &    111.09618    &     28.917578   &    -2.9  $\pm$      4.0   &    -6.9    $\pm$    4.0  &   15.037  &   14.425  &   14.213  &   14.163  &   14.266  &    -212.8   &   0.77  &   dKs7.0   \\
 J072557.18+291137.1     &    111.48827    &     29.193658   &     9.1  $\pm$      3.8   &   -10.1    $\pm$    3.8  &   14.300  &   13.648  &   13.431  &   13.371  &   13.290  &      18.6   &   0.35  &   dMr1.5   \\
 J072106.54+284001.0     &    110.27729    &     28.666959   &    -5.4  $\pm$      3.9   &   -23.0    $\pm$    3.9  &   14.347  &   13.670  &   13.462  &   13.460  &   13.526  &       6.5   &   0.56  &   dKr7.5   \\
 J072037.71+292324.7     &    110.15716    &     29.390220   &     1.1  $\pm$      3.9   &    -3.1    $\pm$    3.9  &   14.565  &   13.855  &   13.617  &   13.561  &   13.495  &       8.5   &   0.43  &   dMp0.5   \\

\hline
\multicolumn{13}{l}{{\sc Notes:} Designation is from the LAMOST DR1; $\mu_{\alpha}$ cos($\delta$) and $\mu_{\delta}$ are poper motions from the PPMXL; J, H, and $K_{\rm s}$ are 2MASS near infrared magnitude; W1 and W2 are WISE}\\
\multicolumn{13}{l}{  infrared magnitude; RV is the radial velocity we measured from the LAMOST spectra; Dist is the spectroscopic distance based on the M$_{\rm j}$ magnitude; SpTy is the spectral subtype} \\
\multicolumn{13}{l}{  classified by our template fit pipeline. The entire table is in its electric form.} \\
\end{tabular}
%\normalsize
%\end{sidewaystable}
\end{table*}

\section{Conclusions and discussion}
\label{sect:conclusion}
We have successfully assembled a set of M giant templates from M0 to M6 by using the LAMOST DR1 spectra. After combining the M giant templates and M dwarf/subdwarf templates as a new M-type spectral library, we re-run the updated classification pipeline to identify and classify M-type stars in the LAMOST DR1. The 8639 M giants and the 101690 M dwarfs/subdwarfs are cataloged. We present the information of celestial coordinates, JHK$_{s}$ infrared magnitudes in 2MASS, spectral subtypes, radial velocity and derived spectroscopic distance.

Based on our M-type stars catalog, Li et al. (2015,in preparation) developed a new photometric method to separate M giants from M dwarfs catalog. The WISE bands are found to be more efficient to separate M giants and dwarfs than the 2MASS bands. Figure~\ref{color} shows the distribution of our M giants and dwarfs catalog in [J-K$_s$,W1-W2] color-color diagram. Two colors represent M giants (red dots) and dwarfs (blue dots) sample which were classified by our spectral classification pipeline. As expected, there are significant difference between the giants and dwarfs in the infrared colors. By using the criteria of mean SNR greater than 5, the M dwarf contamination rate is about 4.7 \% in our giant sample and the M giant contamination rate is about 0.2 \% in our dwarf sample. By increasing the SNR criteria of our sample, the contamination rate will be smaller.

In Figure~\ref{color}, we note that there is a tail in the giants sample, from -0.1 to 0.1 in the (W1-W2)$_{0}$ and 0.9 to 1.3 in the (J-K$_s$)$_{0}$. We carefully examine these tail stars and believe that they are more likely to be metal poor stars ( see more details in Li et al.2015,in preparation).

Although the different locations of M giants and M dwarfs sample in Figure~\ref{color} clearly shows that our classification pipeline separate the two stellar populations more efficiently, a small number of outliers in the giant sample are located on the dwarf region (-0.1 $\leq$ (W1-W2)$_{0}$ $\leq$ 0, 0 $\leq$ (J-K$_s$)$_{0}$ $\leq$ 0.7). After checking these stars by their spectra, we find that the possible contamination are including late K-type dwarfs, early M dwarfs, binaries as well as some low S/N spectra.
\normalem

\begin{acknowledgements}

We thank Guhathakurta Puragra and Katie Hamren for helpful information and comments. This research was supported by `973 Program' 2014 CB845702, the Strategic Priority Research Program "The Emergence of Cosmological Structures" of the Chinese Academy of Sciences, Grant No. XDB09000000,  and the National Science Foundation of China (NSFC) under grants 11173044 (PI:Hou), by the Shanghai Natural Science Foundation 14ZR1446900 (PI:Zhong), by the Key Project 10833005 (PI:Hou), and by the Group Innovation Project NO.11121062.

Guoshoujing Telescope (the Large Sky Area Multi-Object Fiber Spectroscopic Telescope LAMOST) is a National Major Scientific Project built by the Chinese Academy of Sciences. Funding for the project has been provided by the National  Development and Reform Commission. LAMOST is operated and managed by the   National Astronomical Observatories, Chinese Academy of Sciences.

\end{acknowledgements}


\begin{thebibliography}{}

\bibitem[Allen
\& Strom(1995)]{1995AJ....109.1379A} Allen, L.~E., \& Strom, K.~M.\ 1995, \aj, 109, 1379

\bibitem[Bessell \& Brett(1988)]{1988PASP..100.1134B} Bessell, M.~S.,
 \& Brett, J.~M.\ 1988, \pasp, 100, 1134

\bibitem[Bessell et al.(1998)]{1998A&A...333..231B} Bessell, M.~S.,
 Castelli, F., \& Plez, B.\ 1998, \aap, 333, 231

\bibitem[Bochanski et al.(2007)]{2007AJ....133..531B} Bochanski, J.~J.,
West, A.~A., Hawley, S.~L., \& Covey, K.~R.\ 2007, \aj, 133, 531

\bibitem[Bochanski et al.(2014a)]{2014AJ....147...76B} Bochanski, J.~J.,
Willman, B., West, A.~A., Strader, J., \& Chomiuk, L.\ 2014a, \aj, 147, 76

\bibitem[Bochanski et al.(2014b)]{2014ApJ...790L...5B} Bochanski, J.~J.,
Willman, B., Caldwell, N., et al.\ 2014b, \apjl, 790, L5

\bibitem[Chen et al.(2012)]{2012RAA....12..805C} Chen, L., Hou, J.-L.,
  Yu, J.-C., et al.\ 2012, Res. Astron. Astrophys., 12, 805

\bibitem[Covey et al.(2007)]{2007AJ....134.2398C} Covey, K.~R., Ivezi{\'c},
{\v Z}., Schlegel, D., et al.\ 2007, \aj, 134, 2398

\bibitem[Covey et al.(2008)]{2008AJ....136.1778C} Covey, K.~R., Hawley,
S.~L., Bochanski, J.~J., et al.\ 2008, \aj, 136, 1778

\bibitem[Cui et al.(2012)]{Cui.2012} Cui, X.-Q., Zhao, Y.-H.,
Chu, Y.-Q., et al.\ 2012, Research in Astronomy and Astrophysics, 12, 1197

\bibitem[Deng et al.(2012)]{deng12} Deng, L.~C., Newberg, H. J., Liu,
  C., et al.\ 2012, Res. Astron. Astrophys., 12, 735

\bibitem[Danks
\& Dennefeld(1994)]{1994PASP..106..382D} Danks, A.~C., \& Dennefeld, M.\ 1994, \pasp, 106, 382

\bibitem[Fluks et
al.(1994)]{1994A&AS..105..311F} Fluks, M.~A., Plez, B., The, P.~S., et al.\ 1994, \aaps, 105, 311

\bibitem[Lan{\c c}on
\& Wood(2000)]{2000A&AS..146..217L} Lan{\c c}on, A., \& Wood, P.~R.\ 2000, \aaps, 146, 217

\bibitem[Lawrence et al.(2007)]{2007MNRAS.379.1599L} Lawrence, A., Warren,
S.~J., Almaini, O., et al.\ 2007, \mnras, 379, 1599

\bibitem[L\'epine, Rich, \& Shara(2003)]{Lepine.2003}
L\'epine, S., Rich, R. M., \& Shara, M. M. 2003, \aj, 125, 1598

\bibitem[L{\'e}pine et al.(2007)]{2007ApJ...669.1235L} L{\'e}pine, S.,
Rich, R.~M., \& Shara, M.~M.\ 2007, \apj, 669, 1235

\bibitem[L\'epine, Rich, \& Shara(2007)]{Lepine.2007}
L\'epine, S., Rich, R. M., \& Shara, M. M. 2007, \apj, 669, 1235

\bibitem[L{\'e}pine
\& Gaidos(2011)]{2011AJ....142..138L} L{\'e}pine, S., \& Gaidos, E.\ 2011, \aj, 142, 138

\bibitem[L{\'e}pine et al.(2013)]{2013AJ....145..102L} L{\'e}pine, S.,
Hilton, E.~J., Mann, A.~W., et al.\ 2013, \aj, 145, 102

\bibitem[Luo et al.(2012)]{luo12} Luo, A., Zhang, H., Zhao, Y., Zhao,
  G., Cui, X., Li, G., Chu, Y., et al.\ 2012, Res. Astron. Astrophys.,
  12, 1243

\bibitem[Luo et al.(2015a)]{luo15a} Luo A.-L. et al., 2015a, RAA, in press

\bibitem[Luo et al.(2015b)]{luo15b} Luo A.-L. et al., 2015b, RAA, in press

\bibitem[Mann et al.(2012)]{2012ApJ...753...90M} Mann, A.~W., Gaidos, E.,
L{\'e}pine, S., \& Hilton, E.~J.\ 2012, \apj, 753, 90

\bibitem[Majewski et al.(2003)]{2003ApJ...599.1082M} Majewski, S.~R.,
Skrutskie, M.~F., Weinberg, M.~D.,
\& Ostheimer, J.~C.\ 2003, \apj, 599, 1082

\bibitem[Majewski et al.(2011)]{2011ApJ...739...25M} Majewski, S.~R.,
Zasowski, G., \& Nidever, D.~L.\ 2011, \apj, 739, 25

\bibitem[Montes et al.(1999)]{1999ApJS..123..283M} Montes, D., Ramsey,
L.~W., \& Welty, A.~D.\ 1999, \apjs, 123, 283

\bibitem[Reid et al.(1995)]{1995AJ....110.1838R} Reid, I.~N., Hawley,
S.~L., \& Gizis, J.~E.\ 1995, \aj, 110, 1838

\bibitem[Ro\"eser et al.(2010)]{Roeser.2010}
Ro\"eser S., Demleitner M., \& Schilbach E. 2010, AJ, 139, 2440

\bibitem[Skrutskie et al.(2006)]{Skrutskie.2006} Skrutskie, M.~F.,
Cutri, R.~M., Stiening, R., et al.\ 2006, \aj, 131, 1163

\bibitem[Schlegel et al.(1998)]{1998ApJ...500..525S} Schlegel, D.~J.,
Finkbeiner, D.~P., \& Davis, M.\ 1998, \apj, 500, 525

\bibitem[Sharma et al.(2010)]{2010ApJ...722..750S} Sharma, S., Johnston,
K.~V., Majewski, S.~R., et al.\ 2010, \apj, 722, 750

\bibitem[West et al.(2011)]{2011AJ....141...97W} West, A.~A., Morgan,
D.~P., Bochanski, J.~J., et al.\ 2011, \aj, 141, 97

\bibitem[Wright et al.(2010)]{2010AJ....140.1868W} Wright, E.~L.,
Eisenhardt, P.~R.~M., Mainzer, A.~K., et al.\ 2010, \aj, 140, 1868

\bibitem[York et al.(2000)]{2000AJ....120.1579Y} York, D.~G., Adelman, J.,
Anderson, J.~E., Jr., et al.\ 2000, \aj, 120, 1579

\bibitem[Yi et al.(2014)]{2014AJ....147...33Y} Yi, Z., Luo, A., Song, Y.,
et al.\ 2014, \aj, 147, 33

\bibitem[Zhao et al.(2012)]{2012RAA....12..723Z} Zhao, G., Zhao, Y.-H.,
Chu, Y.-Q., Jing, Y.-P., \& Deng, L.-C.\ 2012, Research in Astronomy and Astrophysics, 12, 723

\bibitem[Zhong et al.(2015)]{zhong.2015} Zhong, J. et al. \ 2015, AJ, submitted (arXiv:1505.01592)

\end{thebibliography}
\end{document}